\documentclass[a4paper,10pt,twoside]{cpc-hepnp}
\usepackage{CJK,upgreek,fancyhdr}

\usepackage{subfigure}
\usepackage{multicol}
\usepackage{booktabs}
\usepackage{amsmath}
\usepackage{amsfonts,amssymb,bm,mathrsfs,bbm,amscd}
\usepackage{lastpage}
\usepackage[pdftex]{graphicx}
\usepackage{multirow}
\DeclareGraphicsExtensions{.pdf,.jpeg,.png}
\usepackage{epstopdf}
\usepackage[english]{babel}
\usepackage[colorlinks=false,pdfpagemode=FullScreen,,setpagesize=off,pdfborder={0 0 0}]{hyperref}
\usepackage{overpic}
\usepackage{lineno}
\usepackage{color}
\usepackage{diagbox}
\usepackage{xurl}

\newcommand{\ipb}{\mathrm{pb}^{-1}}

\newcommand{\ee}{e^+e^-}
\newcommand{\pip}{\pi^+}

\newcommand{\mev}{\,\mathrm{MeV}}

\newcommand{\gev}{\,\mathrm{GeV}}
\newcommand{\gevc}{\,\mathrm{GeV}/c}

\begin{document}
\begin{CJK*}{UTF8}{gkai}

\fancyhead[c]{\small Chinese Physics C~~~Vol. xx, No. x (2021) xxxxxx}
\fancyfoot[C]{\small 010201-\thepage}
\footnotetext[0]{Received xxxx June xxxx}

\title{Luminosities and energies of $e^+e^-$ collision data taken between $\sqrt{s}$=4.61 GeV and 4.95 GeV at BESIII\thanks{
Supported in part by National Key R$\&$D Program of China under Contracts Nos. 2020YFA0406400, 2020YFA0406300; National Natural Science Foundation of China (NSFC) under Contracts Nos. 11635010, 11735014, 11805086,  11835012, 11935015, 11935016, 11935018, 11961141012, 12022510, 12025502, 12035009, 12035013, 12192260, 12192261, 12192262, 12192263, 12192264, 12192265; the Chinese Academy of Sciences (CAS) Large-Scale Scientific Facility Program; Joint Large-Scale Scientific Facility Funds of the NSFC and CAS under Contract No. U1832207; CAS Key Research Program of Frontier Sciences under Contract No. QYZDJ-SSW-SLH040; 100 Talents Program of CAS; Fundamental Research Funds for the Central Universities, Lanzhou University, University of Chinese Academy of Sciences; The Institute of Nuclear and Particle Physics (INPAC) and Shanghai Key Laboratory for Particle Physics and Cosmology; ERC under Contract No. 758462; European Union's Horizon 2020 research and innovation programme under Marie Sklodowska-Curie grant agreement under Contract No. 894790; German Research Foundation DFG under Contracts Nos. 443159800, Collaborative Research Center CRC 1044, GRK 2149; Istituto Nazionale di Fisica Nucleare, Italy; Ministry of Development of Turkey under Contract No. DPT2006K-120470; National Science and Technology fund; National Science Research and Innovation Fund (NSRF) via the Program Management Unit for Human Resources $\&$ Institutional Development, Research and Innovation under Contract No. B16F640076; STFC (United Kingdom); Suranaree University of Technology (SUT), Thailand Science Research and Innovation (TSRI), and National Science Research and Innovation Fund (NSRF) under Contract No. 160355; The Royal Society, UK under Contracts Nos. DH140054, DH160214; The Swedish Research Council; U. S. Department of Energy under Contract No. DE-FG02-05ER41374.}}
\maketitle
\begin{center} 
\begin{small}
\begin{center}
M.~Ablikim(麦迪娜)$^{1}$, M.~N.~Achasov$^{11,b}$, P.~Adlarson$^{70}$, M.~Albrecht$^{4}$, R.~Aliberti$^{31}$, A.~Amoroso$^{69A,69C}$, M.~R.~An(安美儒)$^{35}$, Q.~An(安琪)$^{53,66}$, X.~H.~Bai(白旭红)$^{61}$, Y.~Bai(白羽)$^{52}$
% ...
\\
\vspace{0.2cm}
(BESIII Collaboration)\\
\vspace{0.2cm} {\it
$^{1}$ Institute of High Energy Physics, Beijing 100049, People's Republic of China\\
% ...
$^{4}$ Bochum Ruhr-University, D-44780 Bochum, Germany\\
% ...
$^{11}$ G.I. Budker Institute of Nuclear Physics SB RAS (BINP), Novosibirsk 630090, Russia\\
% ...
$^{31}$ Johannes Gutenberg University of Mainz, Johann-Joachim-Becher-Weg 45, D-55099 Mainz, Germany\\
% ...
$^{35}$ Liaoning Normal University, Dalian 116029, People's Republic of China\\
% ...
$^{52}$ Southeast University, Nanjing 211100, People's Republic of China\\
$^{53}$ State Key Laboratory of Particle Detection and Electronics, Beijing 100049, Hefei 230026, People's Republic of China\\
% ...
$^{61}$ University of Jinan, Jinan 250022, People's Republic of China\\
% ...
$^{66}$ University of Science and Technology of China, Hefei 230026, People's Republic of China\\
% ...
$^{69}$ (A)University of Turin, I-10125, Turin, Italy; (B)University of Eastern Piedmont, I-15121, Alessandria, Italy; (C)INFN, I-10125, Turin, Italy\\
$^{70}$ Uppsala University, Box 516, SE-75120 Uppsala, Sweden\\
% ...
$^{b}$ Also at the Novosibirsk State University, Novosibirsk, 630090, Russia\\
% ...
}\end{center}

\vspace{0.4cm}
\end{small}
\end{center}

\begin{abstract}
From December 2019 to June 2021, the BESIII experiment collected  about 5.85 fb$^{-1}$ of data at center-of-mass energies between 4.61~GeV and 4.95~GeV.
This is the highest collision energy BEPCII has reached so far.
The accumulated $e^+e^-$ annihilation data samples are useful for studying charmonium(-like) states and charmed-hadron decays.
By adopting a novel method of analyzing the production of $\Lambda_{c}^{+}\bar{\Lambda}_{c}^{-}$ pairs in $e^+e^-$ annihilation, the center-of-mass energies are measured with a precision of $\sim$0.6$\mev$.
Integrated luminosities are measured with a precision of better than 1\% by analyzing the events of large-angle Bhabha scattering.
These measurements provide important inputs to the analyses based on these data samples.
\end{abstract}



\begin{multicols}{2}

\section{Introduction}
In 2020, BEPCII implemented an energy upgrade project and increased the maximum center-of-mass energy from 4.61~$\gev$ to 4.95~$\gev$. 
During the data-taking years of 2020 and 2021, the BESIII experiment collected $e^+e^{-}$ annihilation data at 12 center-of-mass energy ($E_{\rm cms}$) points between 4.61~$\gev$ and 4.95~$\gev$.
In this energy region, a few charmonium(-like) states can be produced, such as the Y(4630) and Y(4660)~\cite{Brambilla:2019esw,Guo:2017jvc,Chen:2016qju,Olsen:2017bmm,Olsen:2014qna}, which are potential candidates for multi-quark states other than the charmonium states~\cite{Brambilla:2010cs}. 
More strikingly, at 4.68~$\gev$, the BESIII experiment observed the first candidate for a charged hidden-charm tetraquark with strangeness, $Z_{cs}(3985)^{+}$~\cite{BESIII:2020qkh}. Note that charge conjugation is always implied. 
In addition, the $\Lambda_{c}^{+}\bar{\Lambda}_{c}^{-}$ pair-production is open in this energy region. 
This provides many opportunities for precise measurements of the properties of the lightest charmed baryon $\Lambda_{c}^{+}$, with threshold production and quantum coherence of the accumulated $\Lambda_{c}^{+}\bar{\Lambda}_{c}^{-}$ pairs.
In 2014, the BESIII experiment collected 567~$\ipb$ of $e^+e^{-}$ annihilation data at 4.599~$\gev$, which led to many pioneering measurements~\cite{Li:2021iwf,BESIII:2015bjk,BESIII:2015ysy,BESIII:2018ciw,BESIII:2018mug,BESIII:2017kqg,BESIII:2020kap}. 
About ten times more $\Lambda_{c}^{+}\bar{\Lambda}_{c}^{-}$ pair events are expected to be
contained in all data taken above 4.6~$\gev$, which provides great potential to improve our knowledge of the strong and weak interactions in the charm sector~\cite{Cheng:2021qpd}. The $E_{\rm cms}$ and integrated luminosities of these data samples are important inputs for the analyses using these data samples.


In this paper, we present measurements of $E_{\rm cms}$ and integrated luminosities for data samples at various energy points, as listed in Table~\ref{tab:sum}.
The Beam Energy Measurement System (BEMS)~\cite{Abakumova:2011rp}, which was installed in 2008, is designed to precisely measure the beam energy based on the energies of Compton back-scattered photons.
However, the working range of BEMS is below 4~$\gev$ which implies the measurement of $E_{\rm cms}$ for data samples involved in this paper have to be performed offline. 
A novel method of using $e^+e^{-} \rightarrow\Lambda_{c}^{+}\bar{\Lambda}_{c}^{-}$ events is adopted, which was discussed in the energy measurement for the $\psi$(3770) data at BESIII~\cite{method}.
In the luminosity measurement, the Bhabha scattering process $e^+e^{-} \rightarrow (\gamma)~e^+e^-$ is used, benefiting from its clear signature and large production cross section, which allow for a negligible statistical uncertainty and relatively small systematic uncertainty. 
A cross check of the luminosity results is performed by analyzing the di-photon process $e^+e^- \rightarrow (\gamma)~\gamma\gamma$.

\section{The BESIII detector and MC simulations}

The BESIII detector~\cite{BESIII:2009fln} records symmetric $e^+e^-$ collisions 
provided by the BEPCII storage ring~\cite{Yu:2016cof}, which operates at center-of-mass energies ranging from 2.0~$\gev$ to 4.95~$\gev$.
BESIII has collected large data samples in this energy region~\cite{BESIII:2020nme}. The cylindrical core of the BESIII detector covers 93\% of the full solid angle and consists of a helium-based multilayer drift chamber~(MDC), a plastic scintillator time-of-flight
system~(TOF), and a CsI(Tl) electromagnetic calorimeter~(EMC),
which are all enclosed in a superconducting solenoidal magnet
providing a 1.0~T magnetic field. The solenoid is supported by an
octagonal flux-return yoke with resistive plate counter muon
identification modules interleaved with steel. 
The charged-particle momentum resolution at $1~\gevc$ is
$0.5\%$, and the $dE/dx$ resolution is $6\%$ for electrons
from Bhabha scattering. The EMC measures photon energies with a
resolution of $2.5\%$ ($5\%$) at $1~\gev$ in the barrel (end cap)
region. The time resolution in the TOF barrel region is 68~ps, while
that in the end cap region is 60~ps~\cite{etof}.

Simulated samples produced with a {\sc
geant4}-based~\cite{geant4} Monte Carlo (MC) package, which
includes the geometric description of the BESIII detector and the
detector response, are used to determine detection efficiencies
and to estimate backgrounds. The simulation models the beam
energy spread and initial state radiation (ISR) in the $e^+e^-$
annihilations with the generator {\sc
kkmc}~\cite{ref:kkmc}. 
The inclusive MC sample includes the production of the process $\Lambda_{c}^{+}\bar{\Lambda}_{c}^{-}$ 
using the Born cross section line shape measured by BESIII, open charm processes, the ISR production of vector charmonium(-like) states,
and the continuum processes incorporated in {\sc
kkmc}~\cite{ref:kkmc}. The known decay modes are modelled with {\sc
evtgen}~\cite{ref:evtgen} using branching fractions taken from the
Particle Data Group (PDG)~\cite{ParticleDataGroup:2020ssz}, and the remaining unknown charmonium decays
are modelled with {\sc lundcharm}~\cite{ref:lundcharm}. Final state radiation~(FSR) from charged final state particles is incorporated using {\sc
photos}~\cite{photos}.

\section{Measurement of center-of-mass energies}
\label{sec:cms}


In the process $e^+e^-\rightarrow \Lambda_c^{+}\bar{\Lambda}_c^{-}$, 
each $\Lambda_c^{+}$ ($\bar{\Lambda}_c^{-}$) baryon carries half the energy of the $E_{\rm cms}$. Hence, the $E_{\rm cms}$ is obtained from the calibrated beam energy $E_{\rm \Lambda_c}$ using the reconstructed mass of one $\Lambda_c$  with the following equations:
\begin{eqnarray}
\begin{aligned}
&E_{\rm cms} = 2E_{\rm \Lambda_c}, \\
&E_{\rm \Lambda_c}^{2}=E_{\rm 0}^{2} + m_{\Lambda_c}^{2}c^4 - M_{\rm BC}^{2}c^4.
\label{eq:E2}
\end{aligned}
\end{eqnarray}
Here, $E_{\rm 0}$ is the uncalibrated beam energy, with input values of 2306, 2313, 2320, 2330, 2340, 2350, 2370, 2375, 2390, 2420, 2457 and 2473~$\mev$, respectively, for the beam energies of 12 different energy points, and
$m_{\Lambda_c}$ is the known $\Lambda_{c}$ mass of  $2286.46\pm0.14$~$\mev/c^2$~\cite{ParticleDataGroup:2020ssz,BaBar:2005wur}.
The $M_{\rm BC}$ is the fitted peak position of the beam-constrained mass of the $\Lambda_c$ baryon calculated by $M_{\rm BC}c^2=\sqrt{E_{\rm 0}^{2}-{p}^{2}_{\Lambda_c}{c^2}}$, where ${p}_{\Lambda_c}$ is the momentum of the $\Lambda_c$ measured in the center-of-mass system of the $\ee$ collision. 
Essentially, Eq.~(\ref{eq:E2}) is equivalent to $E_{\rm \Lambda_c}^{2}=p_{\Lambda_c}^{2}c^2 + m_{\Lambda_c}^{2}c^4$.
The distributions of $M_{\rm BC}$, instead of $p_{\Lambda_c}$ are fitted, since the $M_{\rm BC}$  has better resolution and its fit quality is more easily controlled.
According to the above analysis, we expect the calibrated beam energy $E_{\rm \Lambda_c}$ to be stable when different $E_{\rm 0}$ are used and that is well verified by both real data and MC simulations.

To perform this measurement, we use the partial reconstruction method and only one $\Lambda_c^{+}$ is reconstructed which the $\Lambda_{c}^{+}\rightarrow p K^{-} \pip$ channel is used because of its relatively large decay rate and low background contamination.
Each charged track must satisfy the following criteria.
The distance of the closest approach of every charged track to the $\ee$ interaction point (IP) is required to be within 10~cm along the beam direction and within 1~cm in the plane perpendicular to the beam direction. The polar angle $\theta$ between the direction of a charged track and that of the positron beam must satisfy $|\!\cos\theta|<0.93$ for an effective measurement in the active volume of the MDC. 
The $dE/dx$ information recorded by the MDC and the time-of-flight information measured by the TOF are combined to calculate particle identification (PID) probabilities for various particle hypotheses.
Tracks are identified as protons if their PID probabilities ($\mathcal{P}$) satisfy $\mathcal{P}(p)>\mathcal{P}(K)$ and $\mathcal{P}(p)>\mathcal{P}(\pi)$, while charged kaons and pions are identified using $\mathcal{P}(K)>\mathcal{P}(\pi)$ and $\mathcal{P}(\pi)>\mathcal{P}(K)$, respectively. 
All $p K^{-} \pip$ combinations in one event are kept for further study.
In the fit to the $M_{\rm BC}$ distributions, the signals are described by the Bukin function~\cite{bukin} and the backgrounds are described by a linear function.
The fit result of the 4680$\mev$ data sample is shown in Fig.~\ref{fig:fit_result}.

In order to validate the analysis method, an input and output (I/O) check based on the inclusive MC simulation is performed. 
The systematic shifts (0.09~$\sim$~0.25~$\mev$) are noticed between the measured beam energies and the true simulated input values mainly due to the ISR effect. 
The shifts at different energy points are taken into account as individual correction factors. 
The final values of the determined $E_{\rm cms}$ are listed in Table~\ref{tab:sum}.

\begin{center}
\includegraphics[width=0.35\paperwidth]{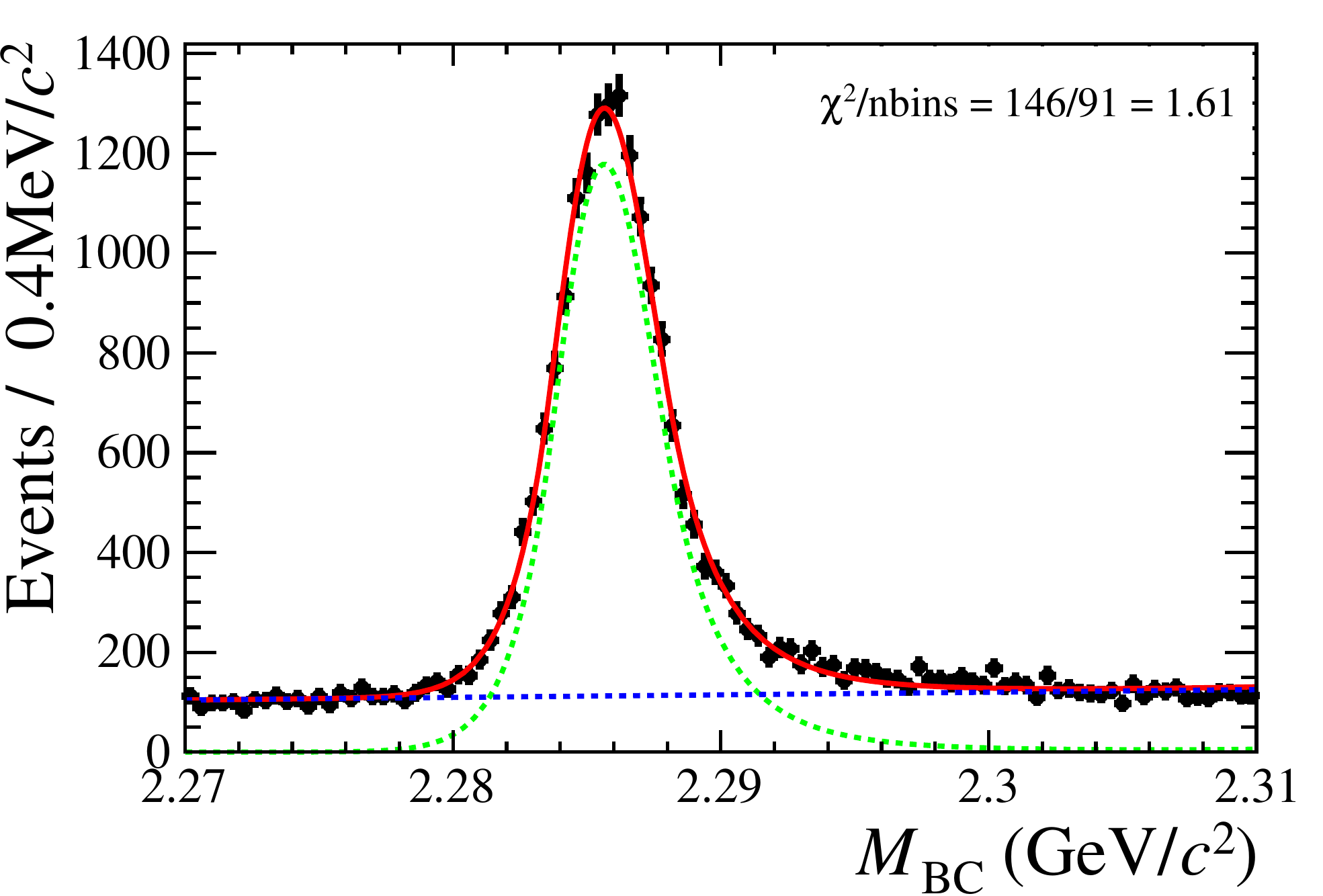}
\figcaption{Fit to the $M_{\rm BC}$ distribution for $\Lambda_{c}^{+}\rightarrow p K^{-} \pip$ candidates from the 4680~$\mev$ data sample.The dotted green line is the fitted signal and the dotted blue line is the fitted background.
Black dots with error bars are data, the red line is the sum of fit functions.}
\label{fig:fit_result}
\end{center}


The systematic uncertainty for the $E_{\rm cms}$ measurement is mainly from the uncertainty of the $\Lambda_{c}$ mass quoted from the PDG, which is 0.28~$\mev$ (twice the uncertainty of the $\Lambda_{c}$ PDG mass).
Other small uncertainties are due to the $M_{\rm BC}$ fit range and the ISR correction.
For the fit range, we vary the fit boundary and repeat the $M_{\rm BC}$ fit.
The maximum relative changes on the  $E_{\rm cms}$ are taken as the systematic uncertainties.

For the ISR correction, we consider the cross section line shape and the influence of the background.
An alternative cross section line shape is first obtained by varying the measured Born cross section line shape of $e^+e^- \to \Lambda_c^+\bar{\Lambda}_c^-$ process within uncertainties.  The alternative and nominal line shape are used to generate signal MC samples of the process $e^+e^- \to \Lambda_{c}^{+}\bar{\Lambda}_{c}^{-}$, where one $\Lambda_c$ decays to $pK\pi$, and the other $\Lambda_c$ decays inclusively, and then the I/O procedure is repeated to get the ISR correction factors. 
The differences in the ISR correction factors are regarded as the systematic uncertainties.
In order to take into account potential effect of background simulation, 
the difference in the ISR correction between the signal MC sample and the inclusive MC sample is regarded as a systematic uncertainty.

For the signal and background shapes, the uncertainties are negligible based on MC simulation studies. 
A summary of systematic uncertainties is given in Table~\ref{tab:sys}. 
For each energy point, the total systematic uncertainty is taken as the quadrature sum of each item.

\begin{table*}[!htbp]
   \caption{Numerical results for the center-of-mass energy $E_{\rm cms}$, the integrated luminosity measured with the Bhabha process $\mathscr{L}_{\rm Bhabha}$, the integrated luminosity measured with the di-photon process $\mathscr{L}_{\rm di-photon}$ and their ratio for all data samples. For the $E_{\rm cms}$ and $\mathscr{L}_{\rm Bhabha}$ measurements, the first uncertainty is statistical and the second is systematic. For the $\mathscr{L}_{\rm di-photon}$ measurement, only statistical uncertainties are presented. For the ratio of $\mathscr{L}_{\rm di-photon}$ 
to $\mathscr{L}_{\rm Bhabha}$ all presented uncertainties are considered.}
   \begin{center}
   \begin{tabular}{ccccc}
	\hline
         \hline
	Sample  &$E_{\rm cms}$ ($\mev$)  &$\mathscr{L}_{\rm Bhabha}$ (pb$^{-1}$) & $\mathscr{L}_{\rm di-photon}$ (pb$^{-1}$)&Ratio ($\%$)\\
	\hline
    4610& 4611.86$\pm$0.12$\pm$0.30&~~103.65$\pm$0.05$\pm$0.55&~103.37$\pm$0.13&~99.73$\pm$0.59\\
	4620& 4628.00$\pm$0.06$\pm$0.32&~~521.53$\pm$0.11$\pm$2.76&~520.17$\pm$0.28&~99.74$\pm$0.55\\
    4640& 4640.91$\pm$0.06$\pm$0.38&~~551.65$\pm$0.12$\pm$2.92&~550.67$\pm$0.29&~99.82$\pm$0.55\\
	4660& 4661.24$\pm$0.06$\pm$0.29&~~529.43$\pm$0.12$\pm$2.81&~527.53$\pm$0.29&~99.64$\pm$0.55\\
    4680& 4681.92$\pm$0.08$\pm$0.29& ~1667.39$\pm$0.21$\pm$8.84&1665.88$\pm$0.51&~99.91$\pm$0.54\\
    4700& 4698.82$\pm$0.10$\pm$0.36&~~535.54$\pm$0.12$\pm$2.84&~533.66$\pm$0.29&99.64$\pm$0.55\\
    4740& 4739.70$\pm$0.20$\pm$0.30&~~163.87$\pm$0.07$\pm$0.87&~165.08$\pm$0.16&100.74$\pm$0.58\\
	4750& 4750.05$\pm$0.12$\pm$0.29&~~366.55$\pm$0.10$\pm$1.94&~367.57$\pm$0.24&~100.28$\pm$0.56\\
	4780& 4780.54$\pm$0.12$\pm$0.30&~~511.47$\pm$0.12$\pm$2.71&~512.03$\pm$0.29&~100.11$\pm$0.55\\
	4840& 4843.07$\pm$0.20$\pm$0.31&~~525.16$\pm$0.12$\pm$2.78&~526.01$\pm$0.30&~100.16$\pm$0.55\\
	4920& 4918.02$\pm$0.34$\pm$0.34&~~207.82$\pm$0.08$\pm$1.10&~208.09$\pm$0.19&~100.13$\pm$0.57\\
	4950& 4950.93$\pm$0.36$\pm$0.38&~~159.28$\pm$0.07$\pm$0.84&~159.85$\pm$0.17&~100.36$\pm$0.58\\

	\hline
	\hline 
   \end{tabular}
    \end{center}
	 \label{tab:sum}
\end{table*}
\begin{table*}[!htbp]
   \caption{Systematic uncertainties for the $E_{\rm cms}$ measurement (in $\mev$). For each energy point, the total systematic uncertainty corresponds to the quadrature sum of each item.}
   \begin{center}
   \begin{tabular}{c|cccccccccccc}
   	\hline
	\hline
	\multirow{2}{*}{Source}& \multicolumn{12}{c}{Sample} \\
	 & 4610 & 4620 & 4640 &4660 &4680 & 4700& 4740 &4750 &4780 &4840 &4920 &4950\\
	\hline
  PDG mass     &0.28&0.28&0.28&0.28&0.28&0.28 &0.28&0.28&0.28&0.28&0.28&0.28\\
 Fit range         &0.04&0.14&0.22&0.04&0.04&0.14&0.04&0.04&0.04&0.02&0.17&0.24\\
 ISR correction   &0.10&0.06&0.13&0.07&0.06&0.17&0.11&0.04&0.09&0.13&0.09&0.08 \\
\hline Total             &0.30&0.32&0.38&0.29&0.29&0.36&0.30&0.29&0.30&0.31&0.34&0.38 \\

	\hline
		\hline
   \end{tabular}
    \end{center}
	 \label{tab:sys}
\end{table*}


We validate the energy measurements to the 12 data samples using the $e^+e^- \rightarrow D^+D^{*-}$ process, where the $D^+ $ is reconstructed via $D^+ \rightarrow K^-\pi^+\pi^+$. The recoil mass of the $D^+$, $RM_{D^+}$, is defined as \begin{eqnarray}
RM_{D^+}=\sqrt{(E_{\rm cms}-E_{D^+})^2/c^4-(\overrightarrow{p}_{\rm cms}-\overrightarrow{p}_{D^+})^2/c^2},
\label{eq:def_mDsRec}
\end{eqnarray}
where $E_{D^+}$ ($\overrightarrow{p}_{D^+}$) is the energy (momentum) of the reconstructed $D^+$. The total energy (momentum) of the initial $\ee$ system, $E_{\rm cms}$ ($\overrightarrow{p}_{\rm cms}$), is input according to our measurement. The peak values of the $RM_{D^+}$ distributions correspond to the known mass of the $D^{*-}$~\cite{ParticleDataGroup:2020ssz}.
To improve the mass resolution~\cite{BESIII:2013mhi}, the variable $RM_{D^+} + M_{D^+}-m_{D^+}$ is adopted to represent the  $D^+$ recoil mass spectrum, where $M_{D^+}$ is the $D^+$ invariant mass, and $m_{D^+}$ is the known $D^+$ mass~\cite{ParticleDataGroup:2020ssz}.
The Bukin function is used to fit the $RM_{D^+} + M_{D^+}-m_{D^+}$ distribution to get the peak mass position, in which  the tail shapes in the signal function are fixed according to MC simulation of the process $e^+e^- \rightarrow D^+D^{*-}$. Following the same procedure used for the ISR correction in the nominal analysis, the measured mass of the $D^{*-}$ in the validation sample shows consistency with the known $D^{*-}$ mass~\cite{ParticleDataGroup:2020ssz}. 
Figure~\ref{Energy_div} shows the mass difference at each energy point, which is consistent with zero and hence validates the measured center-of-mass energies.

\begin{center}
\includegraphics[width=0.35\paperwidth]{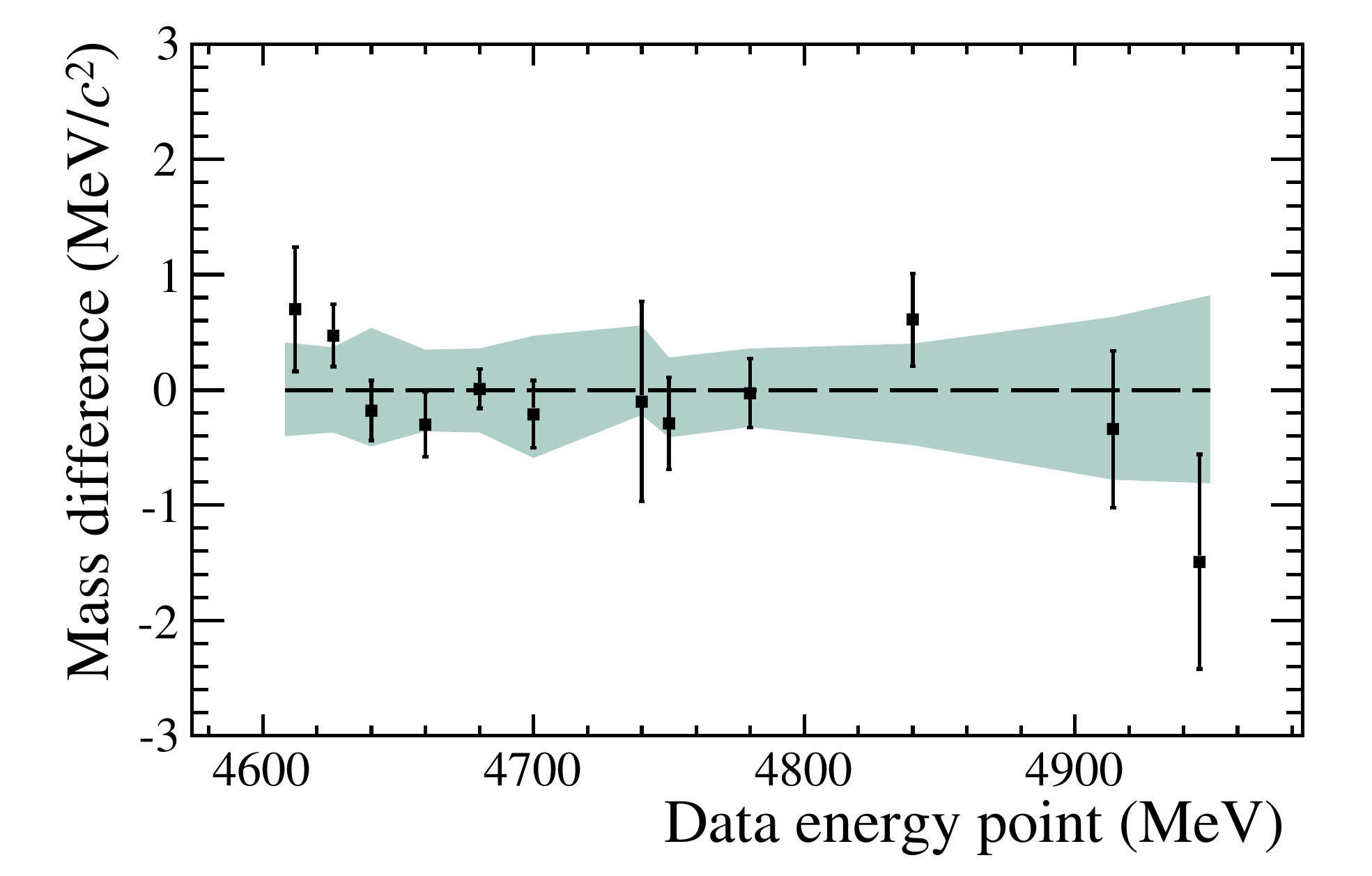}
\figcaption{At each energy point, the difference between the measured $D^{*-}$ mass using the validation sample of $e^+e^- \rightarrow D^+D^{*-}$ and the known $D^{*-}$ mass~\cite{ParticleDataGroup:2020ssz}. Points with error bars are from data and the green band is the uncertainty due to that of the $E_{\rm cms}$ value.}
\label{Energy_div}
\end{center}

\section{Measurement of integrated luminosities}
The integrated luminosity of the data sample is determined by
\begin{eqnarray}\label{Lum_formular}
\mathscr{L} = \frac{N^{\rm obs}_{e^{+}e^{-}\rightarrow X}} { \sigma^{\rm obs}_{e^{+}e^{-}\rightarrow X} \times \epsilon_{e^{+}e^{-}\rightarrow X}},
\end{eqnarray}
where $X$ denotes any specific final state produced in $e^{+}e^{-}$ annihilations, $N^{\rm obs}_{e^{+}e^{-}\rightarrow X}$ is the observed yield for the $e^{+}e^{-}\rightarrow X$ process, $\mathscr{L}$ is the integrated luminosity for data and $\sigma^{\rm obs}_{e^{+}e^{-}\rightarrow X}$ is the visible cross section. 
Here, the Bhabha process ($e^{+}e^{-}\rightarrow (\gamma)e^{+}e^{-}$) is analyzed in the nominal method, and the di-photon $\ee\rightarrow (\gamma)~\gamma\gamma$ process serves as a cross check channel.
The observed cross sections for the two processes are provided by the BabaYaga@NLO generator~\cite{Balossini:2008xr} with 0.1\% precision. 
The configuration parameters for the BabaYaga@NLO generator in generating Bhabha events are listed in Table~\ref{tab:BabaYaga}.

\begin{figure*}[tp]
\begin{center}
\includegraphics[width=0.45\textwidth]{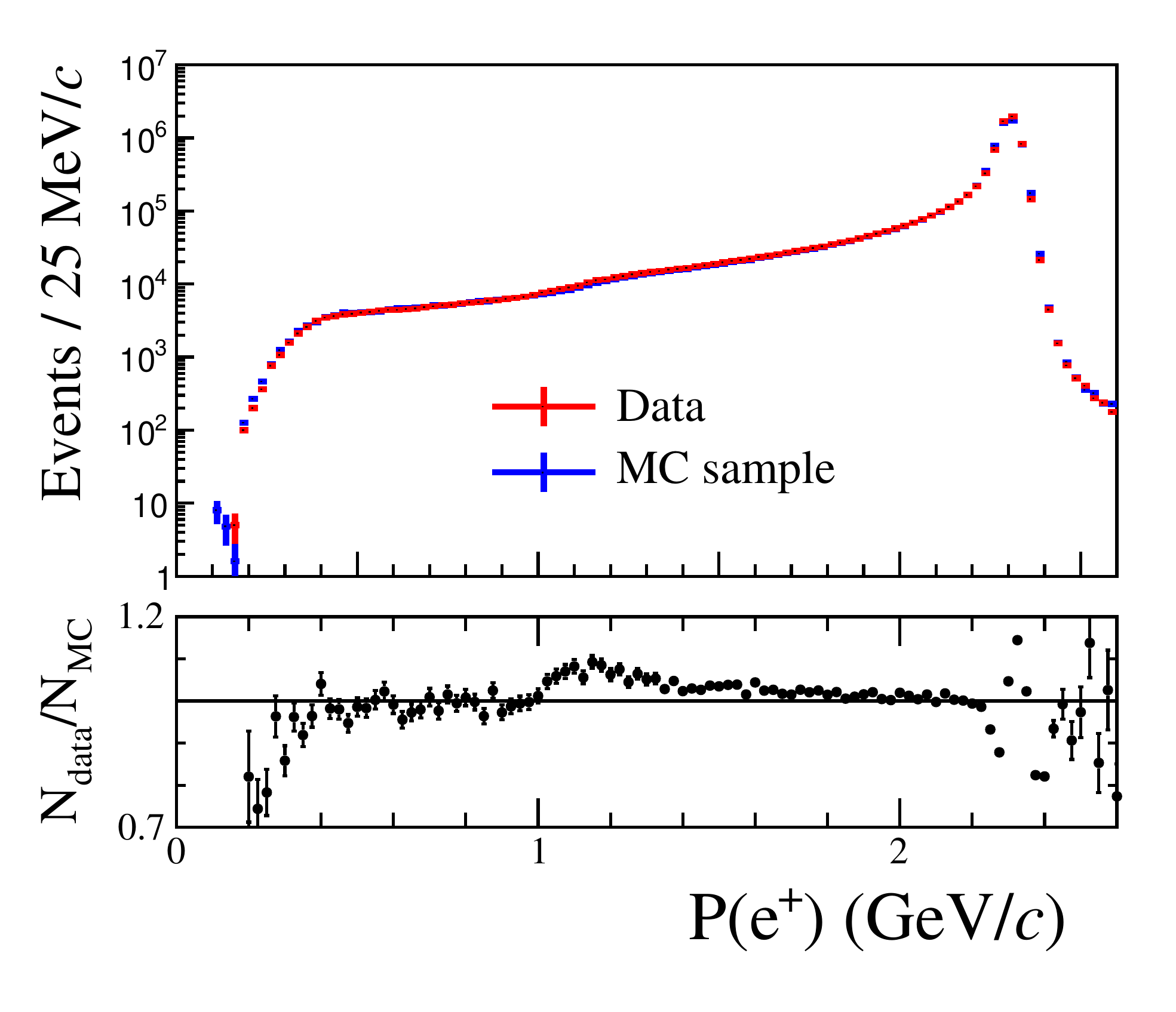}
\includegraphics[width=0.45\textwidth]{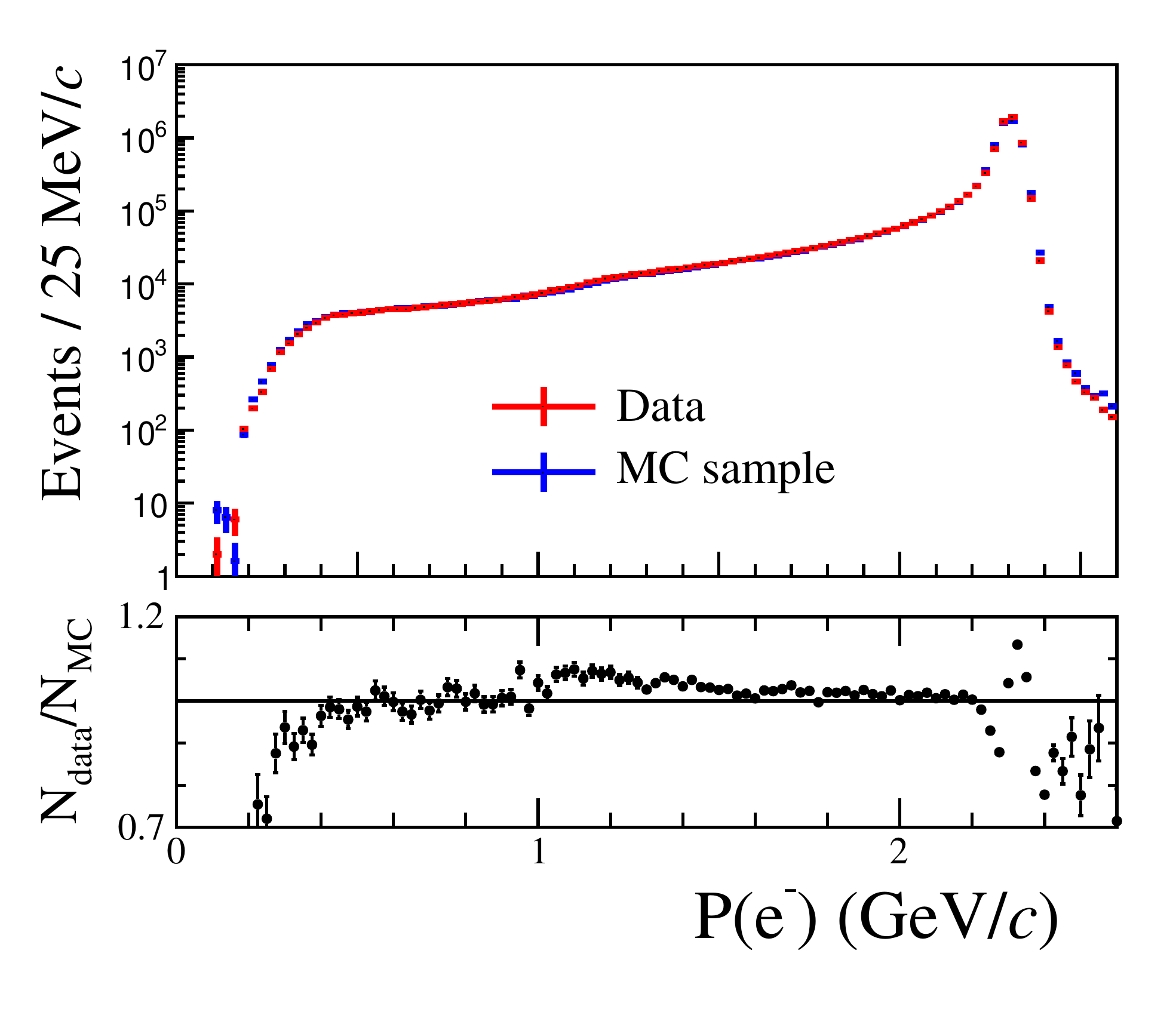}
\includegraphics[width=0.45\textwidth]{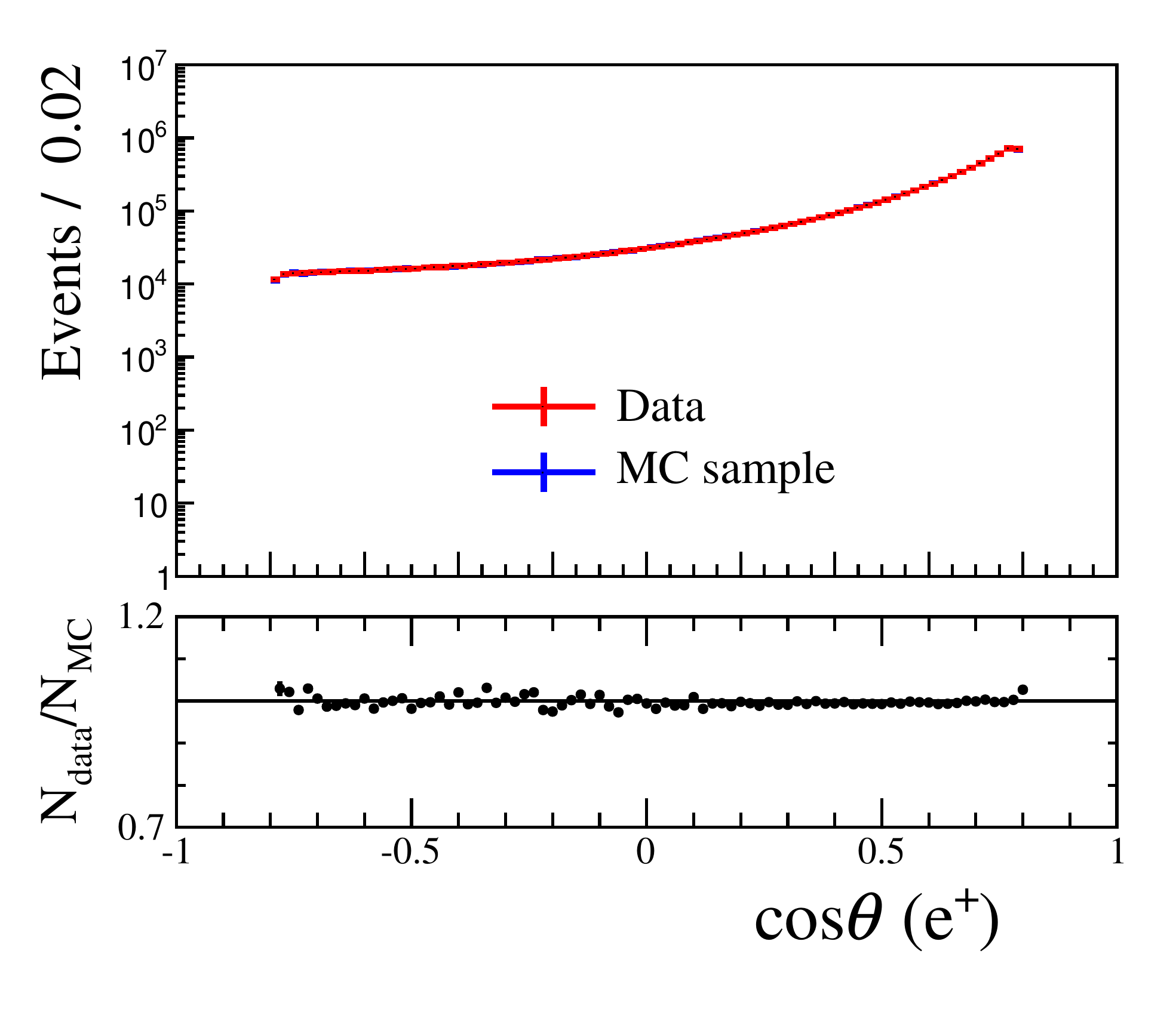}
\includegraphics[width=0.45\textwidth]{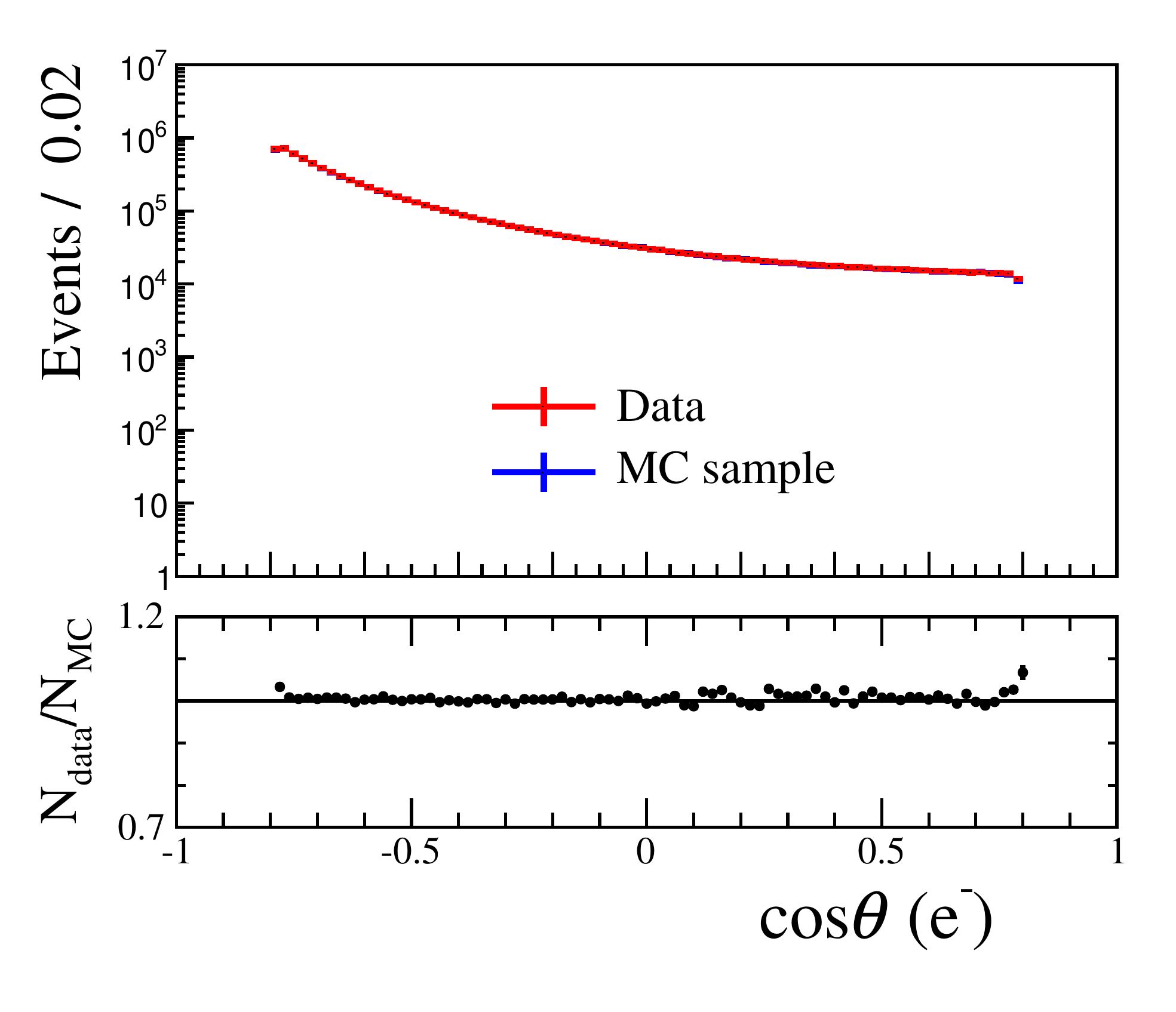}
\includegraphics[width=0.45\textwidth]{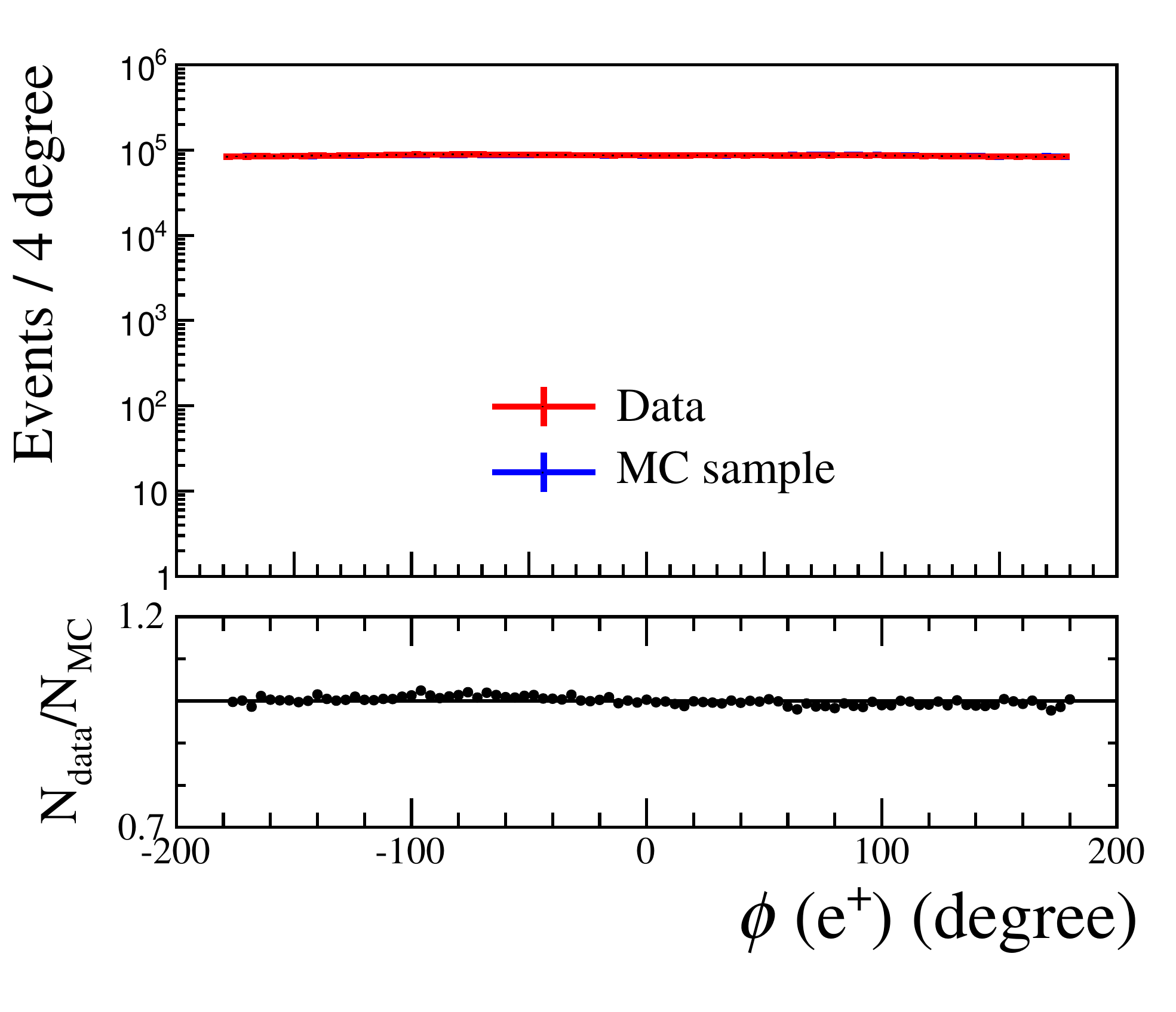}
\includegraphics[width=0.45\textwidth]{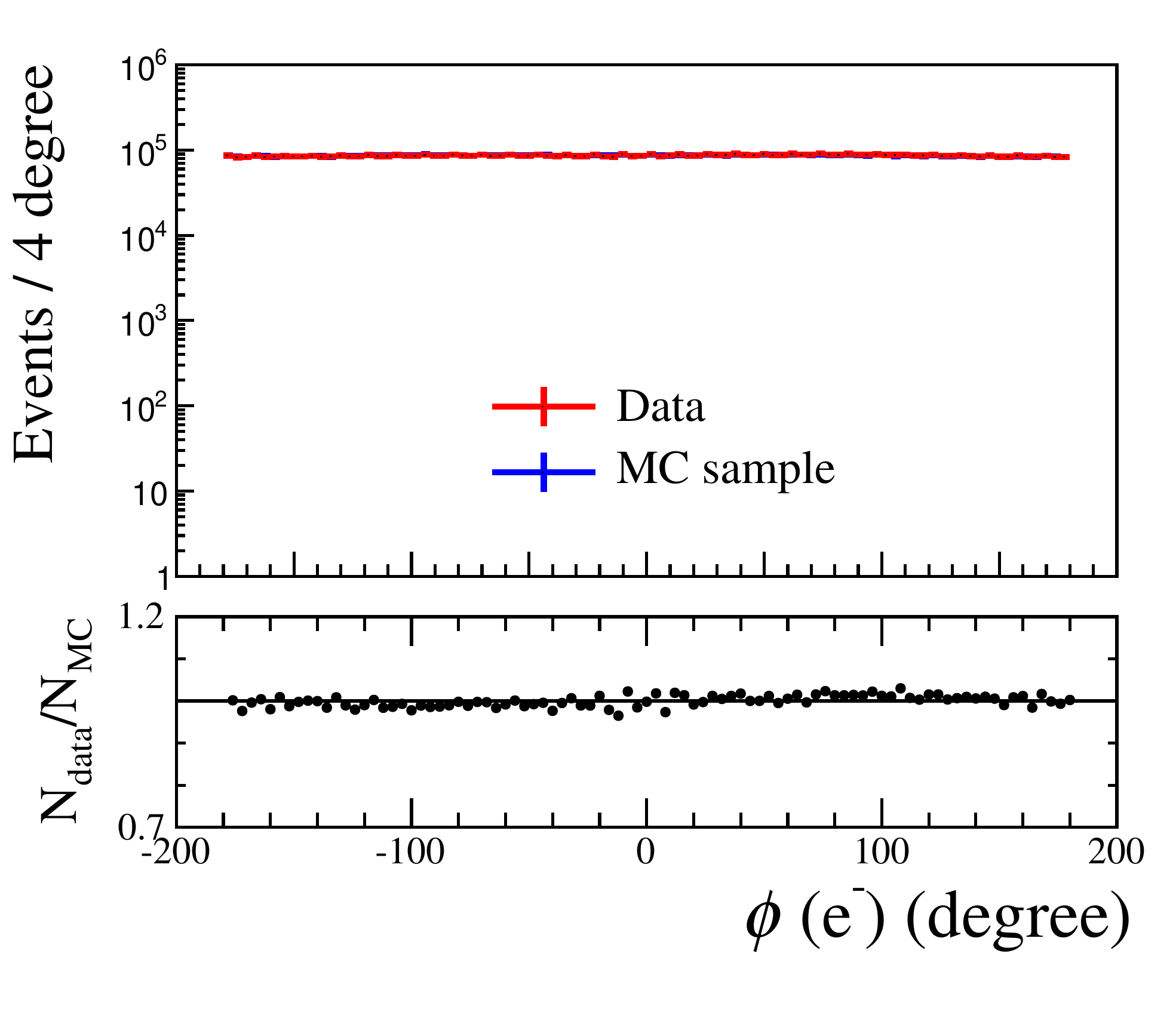}
\figcaption{Comparisons between the data and MC samples at the 4620 $\mev$ energy point for the momentum (top), $\cos\theta$ (middle) and $\phi$ (bottom) distributions for the $e^+$ (left) and $e^-$ (right) in Bhabha events. $N_{data}/N_{MC}$ is the ratio of the data and MC samples. Red points with error bars are data and the blue points are MC samples. The sizes of the MC samples are normalized to those in data.
Except for the variable to be shown, all other requirements used in the event selection have been applied. Although comparison shows inconsistency in high momentum region of the momentum spectrum, this would not have much effect on the final results.}
\label{fig:momentum}
\end{center}
\end{figure*}

The criteria used for selecting Bhabha candidates include the following. 
We require only two oppositely charged tracks (nCharged) detected in the MDC that satisfy $|\!\cos\theta|<$ 0.8 and 
the distance requirement of the closest approach of each charged track to the IP is the same as described in Section 3.
Figure~\ref{fig:momentum} shows the distributions of momentum, polar angle $\cos\theta$ and azimuthal angle $\phi$ for the electron and positron tracks measured in the MDC in data and signal MC samples.
Good consistency between data and MC simulation is shown.
The momentum of each track is required to be larger than 2~$\gevc$ to reject backgrounds from hadronic processes. 
In addition, to suppress the backgrounds from di-photon process, $|\Delta\phi^{\rm EMC}|$ is required to be in the range [5$^{\circ}$, 40$^{\circ}$], where $\Delta\phi^{\rm EMC}$ = $|\phi^{\rm EMC}_1 - \phi^{\rm EMC}_2| -$180$^{\circ}$ and $\phi^{\rm EMC}_{1,2}$ are the azimuthal angles of the two clusters produced by the electron and positron in the EMC in the center-of-mass frame. 



\begin{figure*}[tph]
\centering
\subfigure[]{\includegraphics[width=0.48\textwidth]{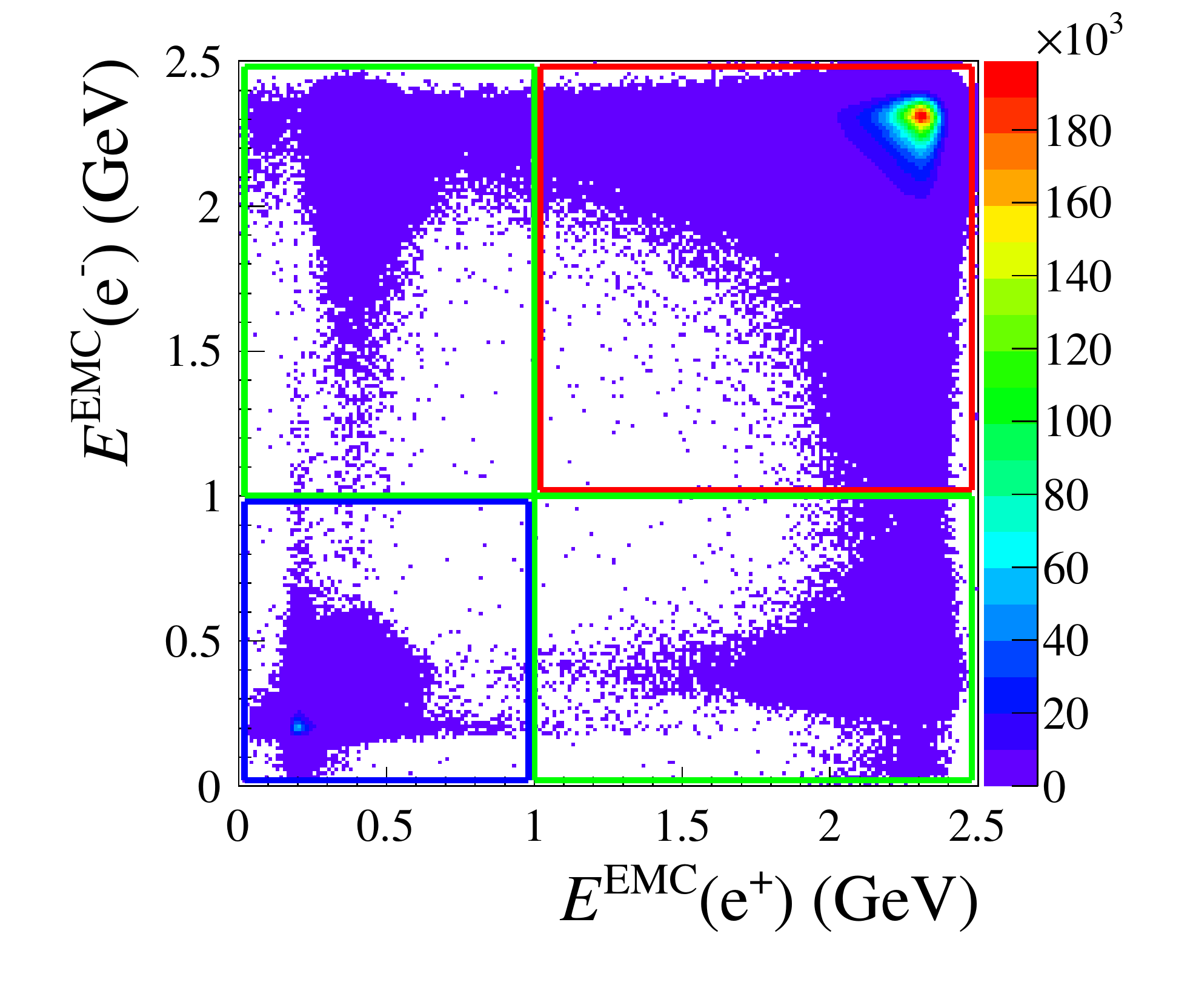}}
\subfigure[]{\includegraphics[width=0.48\textwidth]{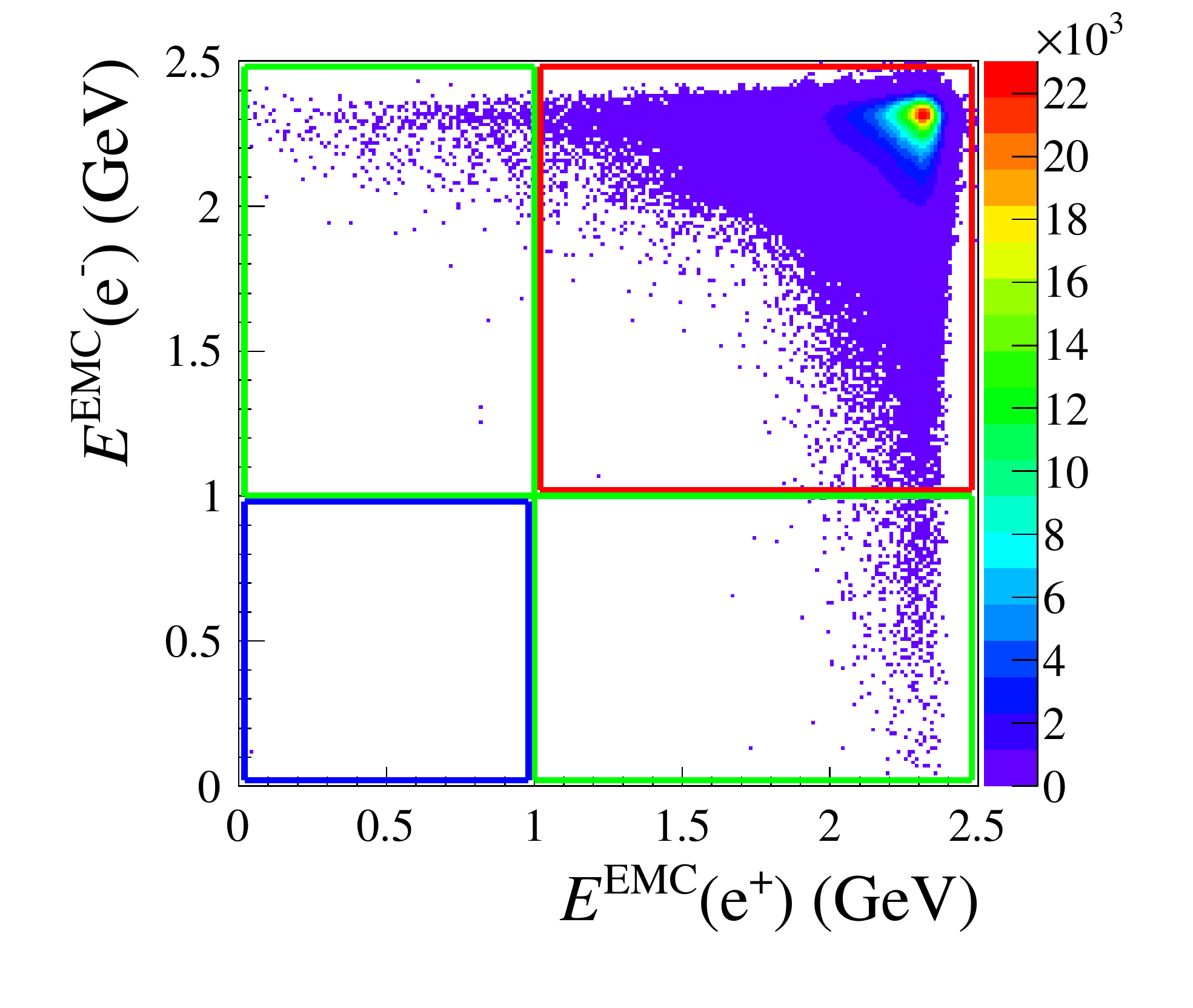}}
\subfigure[]{\includegraphics[width=0.48\textwidth]{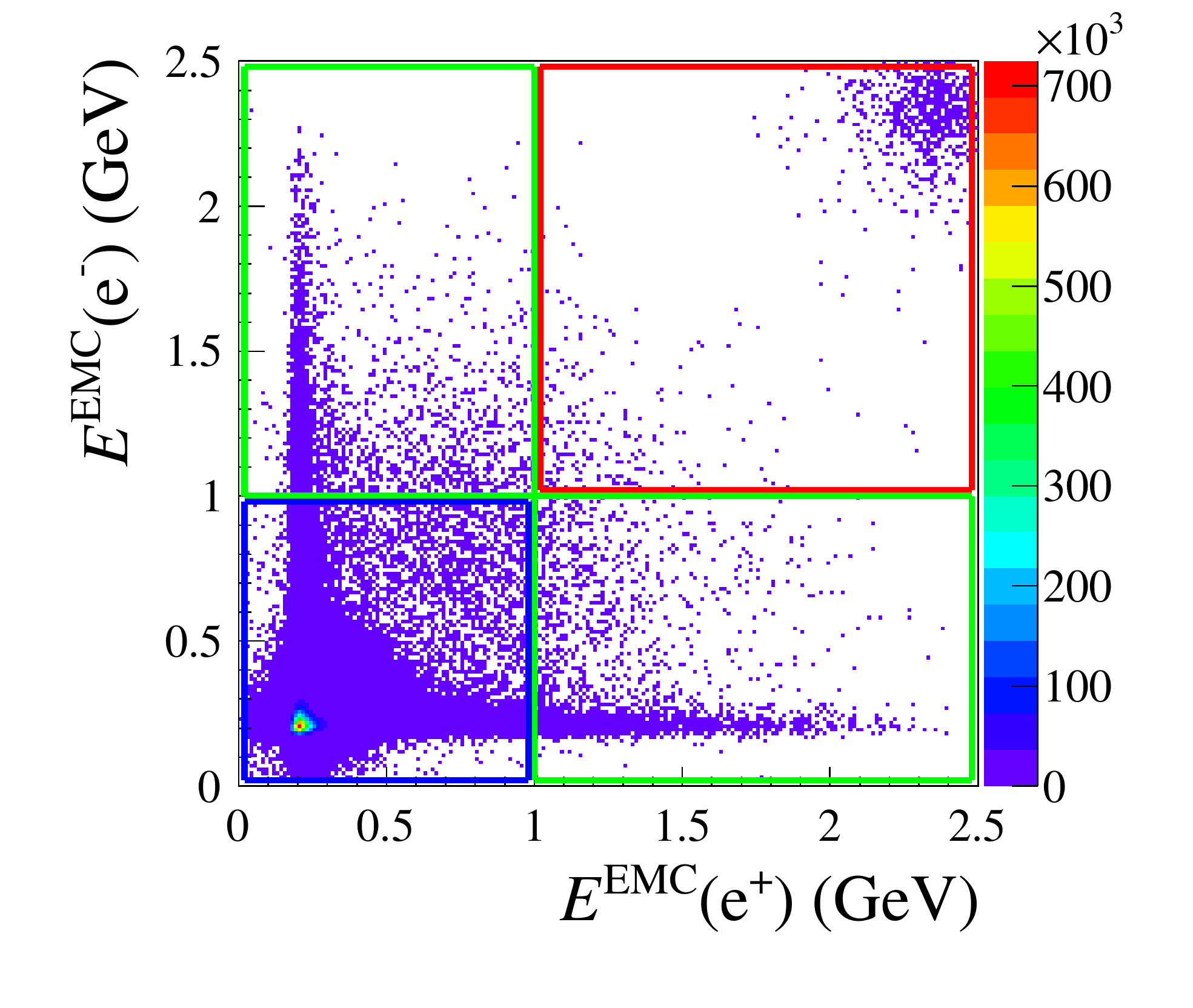}}
\caption{The two-dimensional distributions of $E^{\rm EMC}(e^-)$ versus $E^{\rm EMC}(e^+)$ in data (a),  Bhabha MC sample (b) and background MC samples (c) for the 4620 $\mev$ energy point. 
Three kinematic regions are presented: red square region [$E^{\rm EMC}(e^+)>1~\gev$ and $E^{\rm EMC}(e^-)>1~\gev$] for the \textsc{Normal Sample}, and the remaining regions for the \textsc{Saturation Sample}.
The dimuon backgrounds concentrate in the blue square [$E^{\rm EMC}(e^+)<1~\gev$ and $E^{\rm EMC}(e^-)<1~\gev$], as shown in plot(c).}
\label{fig:2D}
\end{figure*}

Figure~\ref{fig:2D} shows the two-dimensional $E^{\rm EMC}$ distributions of $e^+$ and $e^-$ of Bhabha candidate events, where $E^{\rm EMC}$ is the output of the deposited energies of clusters in the EMC.
Due to the unexpected satuation effect ~\cite{BESIII:2022xii} from a small fraction of EMC electronic readouts, $E^{\rm EMC}$ of the electron and positron becomes underestimated and distributes around 0.4~$\gev$, much less than the expected energies. 
As shown in Fig.~\ref{fig:2D}(a), a fraction  ($2\sim9\%$) of the electron or positron tracks, depending on the track momentum, is influenced by the EMC saturation effect. 
To evaluate the relative size of this effect, the sample is divided into two categories: \textsc{Normal Sample} ($E^{\rm EMC}(e^+)>1~\gev$ and $E^{\rm EMC}(e^-)>1~\gev$) without saturated $\ee$ EMC clusters  and \textsc{Saturation Sample} ($E^{\rm EMC}(e^+)<1~\gev$ or $E^{\rm EMC}(e^-)<1~\gev$) with at least one saturated $e^+$ or $e^-$ EMC cluster. However, as shown in Fig. 4(b), MC simulations do not reflect the saturation effect. In order to obtain the total signal yields correctly matching the MC-determined efficiency, the signal yields in both the \textsc{Normal Sample} and the \textsc{Saturation Sample} are counted in.

For the \textsc{Normal Sample}, 
backgrounds are negligible compared to the statistics of the signal yields, which is validated by using the background MC sample which contains the inclusive MC sample and all QED events except the Bhabha signal, as shown in Fig.~\ref{fig:2D}(c). Hence, the survived \textsc{Normal Sample} is taken as the signal. 
For the \textsc{Saturation Sample}, a portion of the background is from the dimuon process $\ee\to\mu^+\mu^-$, as indicated in Fig.~\ref{fig:2D}(c). To extract the signal yields, the normalized  pulse height, $PH_{\rm norm}$, from the specific ionization energy lost by the charged track in the MDC is adopted to distinguish the Bhabha events from the dimuon backgrounds. Figure~\ref{fig:saturationBKG_fit} shows the fit to the $PH_{\rm norm}$ distribution, where the signals peak around 1.0 for the electron and the dimuon backgrounds peak around 0.86. In the fit, the shape of the electron signals is modelled using the electron sample in the \textsc{Normal Sample}, 
and the muon shape is taken from the control sample of the dimuon process, where the depth of the dimuon tracks penetrating into the muon counter is required to be larger than 10~cm and is implemented for data in the background region ($E^{\rm EMC}(e^+)<1~\gev$ and $E^{\rm EMC}(e^-)<1~\gev$) in Fig.~\ref{fig:2D}(a). 
The fitted yields of the Bhabha process are taken as signals in the \textsc{Saturation Sample}.

The sum of the signal yields in the \textsc{Normal Sample} and \textsc{Saturation Sample} is taken as the total yield of the Bhabha events.
The detection efficiency is estimated  by using the Bhabha MC samples, and the observed Bhabha cross section is calculated based on the BabaYaga@NLO generator. Therefore, the integrated luminosity of the data sample is calculated using Eq.~\eqref{Lum_formular} and the corresponding results for the 12 energy points are given in Table~\ref{tab:sum}. 
The statistical precision of the measured luminosity is better than 0.05$\%$ at each energy point. 

Sources of systematic uncertainties in the luminosity measurement are summarized in Table~\ref{tab:Sys}.
For the 12 energy points, common systematic uncertainties are assigned. Details are discussed as follows.

\begin{center}
   \tabcaption{Configuration of the BabaYaga@NLO generator used for simulating Bhabha events.}
   \begin{tabular*}{80mm}{lc}
	\hline \hline
	Parameter & Value \\
	\hline
      $E_{\rm cms}$ & refer to Table~\ref{tab:sum}\\
      Beam energy spread&1.58~$\mev$\\
      MinThetaAngle&20$^{\circ}$\\
      MaxThetaAngle&160$^{\circ}$\\
      Maximum Acollinearity& 180$^{\circ}$\\
      NSearch & 4000000\\
      RunningAlpha & 1\\
      Number of photon  & $-1$ \\
  	\hline \hline
	 \label{tab:BabaYaga} 
\end{tabular*}
 \end{center}

High momentum electron samples are selected to study the systematic uncertainties due to tracking, the nCharged requirement, the momentum requirement and the $\cos\theta$ requirement. We select one $e^{+}$, with momentum larger than 2~$\gevc$ and $E^{\rm EMC}(e^+)$ greater than 1~$\gev$ as the positron candidate from the Bhabha process, and take the recoil $e^{-}$ as the control sample to study the efficiency of the selection criteria. The relative efficiency difference between the control sample and the Bhabha MC sample is regarded as the systematic uncertainty. 
\begin{center}
\tabcaption{Systematic uncertainties on the luminosity measurement. The total systematic uncertainty corresponds to the quadrature sum of each item.}
   \begin{tabular*}{80mm}{lc}
	\hline 	\hline 
	Source  & Uncertainty ($\%$)\\ \hline
Tracking efficiency&0.30\\
nCharged requirement & 0.10\\
Momentum requirement &0.18\\
 $\cos\theta$ requirement &0.30\\
Saturation events &0.20\\
BabaYaga@NLO generator&0.10\\
MC statistics&0.05\\
Cross section & 0.09\\
\hline
Total & 0.53  \\ \hline
\hline
   \end{tabular*} \label{tab:Sys}
    \end{center}

For estimation of the signal yields, the main systematic issue is in the extraction of the signal yields in  the \textsc{Saturation Sample}. As a check, a different method of counting the number of survived events has been adopted after removing the dimuon backgrounds by discarding events in the region $E^{\rm EMC}(e^+)<1~\gev$ and $E^{\rm EMC}(e^-)<1~\gev$. In the remaining events of the \textsc{Saturation Sample}, there is a small fraction of backgrounds (about 0.50\%), which is neglected. The resultant luminosity differs from the nominal result by 0.20\%, which is taken as the systematic uncertainty.

\begin{center}
\includegraphics[width=0.4\textwidth]{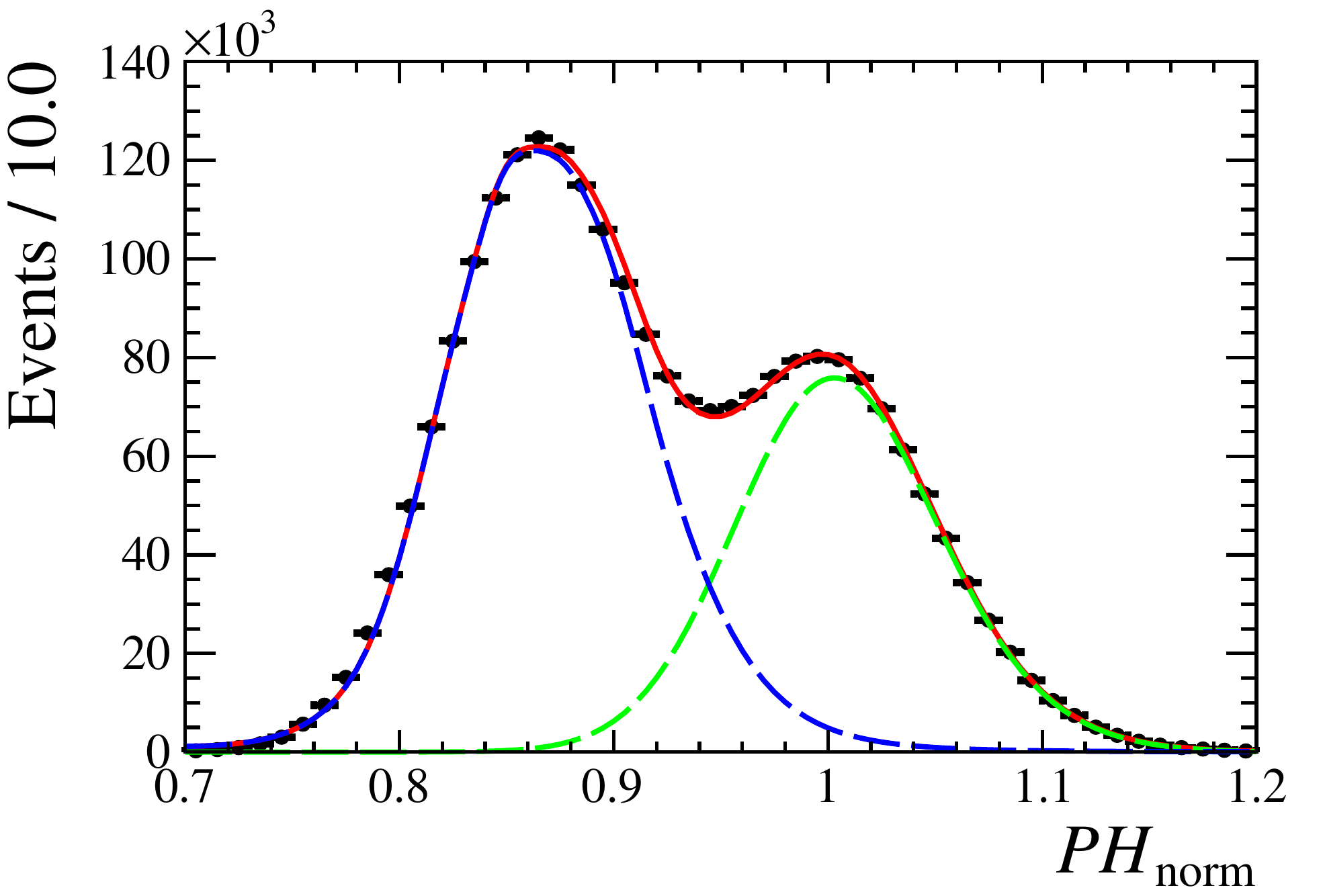}
\figcaption{Fit to the $PH_{\rm norm}$ distribution in the \textsc{Saturation Sample} from the 4620 $\mev$ data sample.  Black dots with error bars are data, the red line is the total fit, the long dashed blue line is background dominated by the dimuon process, and the dashed green line represents saturation events.}
\label{fig:saturationBKG_fit}
\end{center}

For the BabaYaga@NLO Generator, the theoretical uncertainty of the cross section calculation is assigned to be 0.10$\%$~\cite{Balossini:2008xr}.
The systematic uncertainty caused by MC statistics is estimated to be 0.05$\%$ according to the generated 5 million Bhabha MC events for each energy point.

To study the effect due to the $E_{\rm cms}$ uncertainty on the cross section calculation in the BabaYaga@NLO generator, the input values of $E_{\rm cms}$ have been varied within 2~$\mev$ and the corresponding maximum change on the obtained cross section is taken as the systematic uncertainty.

 As a cross check, the di-photon process is used to obtain the luminosity. 
 To select the signal candidates, we require that there are at least two shower clusters in the EMC and no charged tracks detected in the MDC. 
 The clusters must satisfy $|\!\cos\theta^{\rm EMC}|<0.8$.  
 To select back-to-back photon showers and reduce the backgrounds of Bhabha events, their angles with respect to the IP are required to be larger than 178$^{\circ}$ and $\Delta\phi^{\rm EMC}$ of the two showers must be within [$-3^{\circ}$, $3^{\circ}$]. 
To account for the EMC saturation effect and reduce the dimuon backgrounds, the hit number of one EMC shower is required to be larger than 20. The requirements are optimized based on the inclusive MC sample.
The cross section and detection efficiency are determined by the BabaYaga@NLO generator. 
Using Eq.~\eqref{Lum_formular}, the resultant luminosity results at different energy points and the ratios between the measured luminosities based on the Bhabha and di-photon processes are consistent with unity, as given in Table~\ref{tab:sum}. 

\section{Summary}
The center-of-mass energies and the integrated luminosities of the $\ee$ annihilation data  between 4.61~$\gev$ and 4.95~$\gev$ collected from the years 2020 to 2021 with the BESIII detector at the BEPCII collider have been measured with high precision. 
By adopting a novel method for analyzing $\Lambda_{c}^{+}\bar{\Lambda}_{c}^{-}$ pair-production in electron-position annihilations, the center-of-mass energies are measured with a precision of $\sim$0.6~$\mev$, which is dominated by the precision of the known $\Lambda_c$ mass. 
The integrated luminosities of the collected data samples are measured with a precision better than 1\% by analyzing large-angle Bhabha scattering events, after taking into account the EMC saturation effect. These results offer fundamental inputs for physics analyses based on these data samples.

\acknowledgments{
The BESIII collaboration thanks the staff of BEPCII and the IHEP computing center for their strong support. 
}

\end{multicols}

\vspace{-1mm}
\centerline{\rule{80mm}{0.1pt}}
\vspace{2mm}

\begin{multicols}{2}

\end{multicols}

\clearpage
\end{CJK*}
\end{document}